\documentclass[12pt]{article} 

\usepackage{epsfig}
\usepackage{graphicx}
\usepackage{amsmath}
\usepackage{amssymb}
\usepackage{url}
\usepackage{bm}
\newcommand{\ihat}{\hat{\textbf{\i}}}
\newcommand{\jhat}{\hat{\textbf{\j}}}
\newcommand{\khat}{\hat{\mathbf{k}}}

\begin{document}
%
\title{Anomalous accelerations in spacecraft flybys of the Earth}


\author{L. Acedo\thanks{E-mail: luiacrod@imm.upv.es}\\ 
Instituto Universitario de Matem\'atica Multidisciplinar \\
Building 8G, Access C, $2^{\mathrm{o}}$ Floor, Camino de Vera,  \\
Universitat Polit$\grave{\mbox{e}}$cnica de Val$\grave{\mbox{e}}$ncia, 46022, Valencia, Spain}

\maketitle

\begin{abstract}
The flyby anomaly is a persistent riddle in astrodynamics. Orbital analysis in several flybys of the Earth
since the Galileo spacecraft flyby of the Earth in 1990 have shown that the asymptotic post-encounter velocity
exhibits a difference with the initial velocity that cannot be attributed to conventional effects. To  
elucidate its origin, we have developed an orbital program for analyzing the trajectory of the spacecraft in the vicinity
of the perigee, including both the Sun and the Moon's tidal perturbations and the geopotential zonal, tesseral
and sectorial harmonics provided by the EGM96 model. The magnitude and direction of the anomalous acceleration acting upon the spacecraft can be estimated from the orbital determination program by comparing with the trajectories fitted to telemetry data as provided by the mission teams.
This acceleration amounts to a fraction of a mm$/$s$^2$ and
decays very fast with altitude. The possibility of some new physics of gravity in the altitude range for spacecraft flybys
is discussed.
\end{abstract}


%
\section{Introduction}
\label{intro}
The dawn of modern science is deeply intertwined with the improvement in the accuracy of astronomical 
observations. Moreover, in the development of General Relativity \footnote{For recent overviews pointing to unsolved questions and perspectives see \cite{IorioEdUniverse,Debono}} a major role was played by the so-called anomalous
advance of Mercury's perihelion as determined by Urbain Le Verrier in 1859 \cite{LeVerrier,Roseveare}. Later on, Newcomb gave an improved value which it is very close to the one accepted today \cite{Newcomb}. Advances in astronomical instrumentation have also been very important in the subsequent tests of the theory from the determination of the angle of light bending for
rays grazing the Sun in eclipses \cite{Dyson} \footnote{For some recent critical discussions on its implications see  \cite{Kennefick,Longair,Will2015}} to the latest verification of the frame-dragging and geodetical effects on the
Gravity Probe B experiment \cite{Everitt} \footnote{For a comparative discussion of this measurement with other
ones, see \cite{IorioEPL2011}. Overviews of frame-dragging tests can be found in, e.g., \cite{IorioASS2011,Renzetti2013}}. 

In the last fifty years there have also been many developments in the high-accuracy tracking of moons, planets and
spacecraft by means of Doppler effect of radio signals and laser ranging techniques \cite{Dickey1994,Iess2007,Iess2009,Iess2014}. We can fairly say that we are now at an era of high-precision astronomy and astrodynamics in which the level of accuracy for the position and velocity
determination of spacecraft and celestial bodies is several order of magnitude better than it was at the time of 
optical astronomy \cite{Williams1996,Williams2004}. Spacecraft missions are of particular interest for testing gravitational and orbital models and they also provide an opportunity for analyzing trajectories not usually found in natural objects. Moreover, the careful design of spacecraft allows for a better analysis of the physical effects acting upon them in comparison with celestial bodies whose composition and physical parameters are not so well-known. The improvement of measurement techniques has allowed for the discovery of some possible astrometric and gravitational anomalies in recent years such as: anomalous perihelion precessions of the planets, an unexplained secular increase of the eccentricity of the Moon's orbit, the faint young Sun paradox, the secular increase of the mass parameter of the Sun, among other. For comprehensive reviews see  \cite{IorioIJMPDReview,Anderson2010Review}.

Also, the improved high-accuracy tracking of spacecraft have revealed some unexpected phenomena. The recent, but canonical example, of the discovery of an unexplained residue in the tracking analysis of a spacecraft corresponds to the Pioneer anomaly, i. e. , the anomalous constant drift of the redshifted signal received from
the Pioneer 10 and Pioneer 11 spacecraft \cite{Anderson1998,Anderson2002,TuryshevReview}. These spacecraft were sending downlink signals in response to received uplinks since
their launch in the early 1970s until the early 2000s when they were at 80 AU from the Sun and during this time it became apparent that they were exhibiting and anomalous acceleration towards the Sun with magnitude $a_P=(8.74\pm 1.33) \times 10^{-8}$ cm$/$s$^2$ \cite{PioneerPRL}. Despite some hasty claims for new physics as the explanation of this anomaly, it has now been generally accepted that
the origin of the extra acceleration comes from the anisotropic emission of thermal radiation off the spacecraft \cite{PioneerPRL,Rievers2011}. This was already proposed as early as 1998 by Murphy \cite{Murphy1998}. The spacecraft heat is diffusing from the radioisotope thermoelectric generators filled
with Plutonium 238, whose half-life is $87.74$ years, and the decay of this isotope correlates with the diminishing anomalous
acceleration unveiling the classical origin of the Pioneer anomaly. Notwithstanding this explanation in terms
of thermal effects, there are still some researchers who are seeking for some gravitational mechanism to provide a basis for
the origin of the phenomenon (see, for example, the work of Nyambuya \cite{Nyambuya2017}). However, this approach faces the problem of the absence of any noticeable effects, from a similar acceleration, on the orbits of the major bodies of the Solar system, as it has been crucially pointed out by Iorio and Giudice \cite{IorioPioneer2006} and Standish \cite{Standish2008,Standish2010}.

Another lingering anomaly concerns the tidal models for the evolution of the Earth-Moon system. These calculations have disclosed an anomalous increase in the eccentricity of the orbit of the Moon with a value $d e/d t \simeq
(9 \pm 3) \times 10^{-12}$ per year \cite{Williams2001,Williams2003,Williams2014}. This anomaly has not been totally explained
with improved models, although the discrepancy among the models and the observations has been slightly reduced to the presently
accepted value of $d e/ dt \simeq 3 \times 10^{-12}$ yr$^{-1}$ \cite{Williams2016}. Explanations in terms of modified gravity
models have been proposed by Iorio \cite{Iorio2011MNRAS,Iorio2011AJ,Iorio2014Galaxies}.

Another surprising result was discovered during the orbital analysis of the first flyby of the Earth performed by the Galileo
spacecraft on December 8th, 1990 with a perigee of only $960$ km. This analysis revealed a noticeable difference among the
post-encounter and pre-encounter asymptotic velocities of $3.92$ mm$/$s \cite{Anderson2008,LPDSolarSystem}. Subsequent flybys
throughout the years have also shown similar unexplained residues in the fitting of the Doppler tracking data into a single
orbital model. In particular, in the second Galileo flyby (performed on December, 8th, 1992) a total residual velocity decrease of $-8$ mm$/$s was found. In this case, the altitude at perigee was attained inside the thermosphere and atmospheric friction must
be taken into account. However, it has been estimated that only $-3.4$ mm$/$s should correspond to this atmospheric
friction effect \cite{Anderson2008,Acedo2017one}. Other flybys performed by the NEAR, Cassini and Rosetta spacecraft have also exhibited these flyby anomalies.
Anyhow, no anomalies (or negligible ones within the threshold of measurement errors) were found in the Messenger flyby \cite{Anderson2008}, the second and third Rosetta flybys \cite{Jouannic} and, also, in the Juno flyby of October, 9th, 2013 \cite{Thompson}.

This anomaly is still puzzling because no satisfactory conceptual framework for predicting its outcome on the next flyby
has still been found despite ongoing research in the problem.

Anderson et al. \cite{Anderson2008} provided a preliminary phenomenological approach to the data in their seminal work. 
These authors proposed a formula to fit the results for the anomalous asymptotic velocity variation of six flybys of the Earth performed by the NEAR, Galileo (two flybys), Cassini, Rosetta and Messenger between December 1990 and August 2005. 
In this work it was found that the anomaly, $\Delta V_\infty$, could be related to the cosine of the directions defined by the incoming and the outgoing velocity vectors, i.e., the angle of these vectors with the rotation axis of the Earth:

\begin{equation}
\label{AndersonF}
\Delta V_\infty = V_\infty K \left( \cos \delta_i-\cos \delta_o \right)\; ,
\end{equation}

where $V_\infty$ is the asymptotic velocity for the osculating orbit at perigee. Anderson et al. \cite{Anderson2008} speculated that $K$ is a constant related to the quotient of the tangential velocity of the Earth at the Equator and the speed of light as follows:

\begin{equation}
\label{Kpar}
K=\displaystyle\frac{2 \, \Omega \, R_E}{c}=3.099 \times 10^{-6} \; .
\end{equation}

Here $\Omega=7.292115 \times 10^{-5}$ s$^{-1}$ is the angular velocity for the Earth's rotation around its axis, $R_E=6371$ km is the average Earth's radius and $c$ is the speed of light in vacuum. Although, this phenomenological formula provides a good agreement
with the observations of the six flybys analyzed in their paper \cite{Anderson2008}, it has proven to be incorrect in the analysis of subsequent flybys as it does not predict the null results for the asymptotic velocity anomaly obtained for the  Rosetta II and III \cite{Jouannic}, and the most recent Juno flyby of the Earth in October, 2013 \cite{Thompson,IorioJuno}. Moreover, Anderson et al. \cite{Anderson2008} do not provide any explanation of their formula apart from suggesting a connection with an enhanced Lense-Thirring effect not predicted by General Relativity.

We also notice that for the case of Jupiter, and other giant planets, the ratio $K$ in Eq. (\ref{Kpar}) is larger because these bodies spin faster than the Earth and they have also much larger radiuses. For example, for the case of Jupiter we
find $K \simeq 4.19 \times 10^{-5}$. If the idea behind Anderson's formula \cite{Anderson2008} has an element of truth, we can
expect a larger flyby anomaly effect for a spacecraft flyby of Jupiter and anomalous accelerations ten times larger than those
acting in the case of the Earth.

As such a claim as the failure of well-established theories to explain a phenomenon requires a meticulous analysis of all
the possible sources of error or overlooked conventional effects, there have been a sustained effort to evaluate the impact
of classical sources of perturbations in close flybys of the Earth \cite{LPDSolarSystem}. Some of the effects that have been
studied in detail and dismissed as possible explanations (because their impact is small or negligible in comparison with the anomalies) are ocean tides \cite{AcedoMNRAS}, the Lense-Thirring effect \cite{IorioSRE2009,Hackmann}, time-dependent coupling with the tesseral harmonics \cite{AcedoMNRAS},  Lorentz's charge acceleration \cite{Atchison} and thermal radiation \cite{Rievers2011}, among other \cite{LPDSolarSystem}.

An alternative explanation should look at a new kind of interaction including a fifth force or new effects arising in 
extensions of General Relativity. Some of these possibilities have been explored by several authors: Adler studied the
interactions of spacecraft with a putative halo of dark matter particles around the Earth \cite{Adler2010,Adler2011}. Gravity models with retardation effects where suggested by Hafele \cite{Hafele}. Later on, Bel and Acedo pursued this idea by studying an 
extension of Whitehead's theory \cite{Acedo2015,Acedo2017three} and Pinheiro has provided a topological torsion current approach
to try to understand the positive, negative or null values of the anomaly in the different flybys \cite{Pinheiro2014,Pinheiro2016}. Other radical proposals have been given: a modification of inertia  
\cite{McCulloch}, light speed anisotropy \cite{Cahill}, and other non-standard gravity models \cite{Nyambuya2008,Lewis2009, Varieschi2014,Wilhelm2015,Bertolami2016}.

On the present status of this problem, the issue of the conventional or unconventional origin of the flyby anomalies is not
to be solved by theoretical undertakings alone. Obtaining new data and analyzing them properly in previous and
future missions would be the only way to confirm the existence of this anomaly and to qualify it as such, once that all the possible
sources of conventional effects have been taken into account. An excellent opportunity for such a detailed analysis would have been possible with the STE-QUEST mission \cite{STEQUEST} which had been programmed to perform flybys of the Earth at different
altitudes during the successive orbits. With this mission now cancelled, we have to rely on the data from previous or future flybys of the Earth, which are routinely planned on many missions to the outer Solar system.

The objective of this paper is to determine any residuals accelerations remaining in the orbital modelling of several
spacecraft that have performed flybys of the Earth since the Galileo flyby of the Earth in 1990 until the Juno flyby of
2013. To this aim, we will incorporate all the relevant perturbations for a period of an hour before and after the perigee:
tidal perturbations by the Sun and the Moon, atmospheric friction and the geopotential model for the zonal, tesseral and sectorial
contributions to the potential. Moreover, we evaluate the impact of the sources of error in the calculation including mismodelling of the zonal, tesseral and sectorial harmonic coefficients, the effect of Jupiter and other planets in the Solar system, uncertainties in the geocentric latitude, longitude and the obliquity of the ecliptic as well as numerical errors in the integration method. We find that for altitudes below $3000$ km over the surface of the Earth there are statistically significant 
anomalies in the acceleration acting upon the spacecraft of the order of magnitude of $0.1$ mm$/$s$^2$ as expected in 
some models of the flyby anomaly \cite{Acedo2015}. The method described in this paper could allow for the spatial and temporal
resolution of the anomalous force field and we show that the radial, polar and azimuthal components can be estimated.

To this aim we compare the predictions of our model with the ephemeris provided by the mission teams which incorporate the information of telemetry's monitoring to fit the trajectories \cite{Horizons,Giorgini}.

The paper is organized as follows: In sec. \ref{methods} we describe the orbital model and the error analysis procedures
for the integration method. Results on the discrepancies on position and the anomalous acceleration for several flybys are
given in sec. \ref{results}. The paper ends with some conclusions and guidelines for future work in sec. \ref{conclusions}.

\section{Development of the orbital model}
\label{methods}
We have retrieved the trajectories in their Earth's flybys for the following six missions: NEAR (January 1998), Galileo (first flyby performed in December 1990, second flyby in December 1992), Cassini (August 1999) and Juno (October 2013). 
In other flybys, such as the Rosetta flyby of March 2005 or the Messenger of August 2005, the results were either negligible or not statistically significant so we have excluded them from our final analysis. The position and velocity coordinates were obtained from the Horizons web application 
in the ICRF/J2000.0 reference frame with the center of the Earth as origin. The values are obtained with double precision and Doppler ranging data can be considered precise on the range of 
a few cms, especially around the closest approach of the spacecraft to the Earth. We must take into account that improved ephemeris have developed over the years (in the considered period from 1990 to 2013) but the additional bodies incorporated into the new ephemeris \cite{DE431} should not have a relevant impact for the particular section of the trajectory we are interested in this paper.

Another important issue is that the ephemeris for the spacecraft incorporated into Horizons are provided by the mission teams
and they are fits to the telemetry data within the context of the specific orbital model \cite{Giorgini}. In that sense, they can be seen as reflecting the real Doppler and ranging data. So,  whenever we refer in this paper to the data, we are meaning the processed mission data to provide these ephemerides. In contrast with these ephemeris, that take into account the real tracking data, we are using, as comparison, an orbital model that considers all the basic terms of importance for the dynamics close to the perigee.

As our reference initial condition we have chosen the position, ${\bf r}_P$, and velocity, ${\bf v}_P$, of the spacecraft at the 
discrete time instant (in minutes) of the data file corresponding to the closest approach to the Earth. This should not correspond to the true perigee but this is irrelevant for our purpose because our concern is to analyze the trajectory near the Earth were we expect that any anomalous forces giving rise to the flyby anomaly would manifest themselves.

The orbital model in the vicinity of the perigee is obtained by integrating the system of equations of motion for the
position, ${\bf r}$, and velocity, ${\bf v}$, of the spacecraft:

\begin{eqnarray}
\label{eqmotionr}
\displaystyle\frac{d {\bf r}}{d t}&=&{\bf v} \; ,\\
\noalign{\smallskip}
\label{eqmotionv}
\displaystyle\frac{d {\bf v}}{d t}&=&-\mu_E \displaystyle\frac{{\bf r}}{r^3}+\bm{\mathcal{F}}_{\mbox{tidal}}+\bm{\mathcal{F}}_{\mbox{geo}} \; ,
\end{eqnarray}

where $\mu_E=G M_E= 398600.435436$ km$^3/$s$^2$ is the mass constant for the Earth and, apart from the Newtonian monopole, we take into account the tidal, $\bm{{\mathcal F}}_{\mbox{tidal}}$, and geopotential, $\bm{{\mathcal F}}_{\mbox{geo}}$, perturbations in the terms for the total spacecraft acceleration. The tidal acceleration imparted upon the spacecraft located at ${\bf r}$ from a celestial body at
${\bf R}$ is given by:

\begin{equation}
\label{Ftid}
\bm{\mathcal{F}}_{\mbox{tidal}}=\mu \left(-\displaystyle\frac{{\bf R}}{R^3}+\displaystyle\frac{{\bf R}-{\bf r}}{\left(r^2+R^2-2 {\bf r} 
\cdot {\bf R} \right)^{3/2}}\right)\; ,
\end{equation}

where $\mu$ is the mass constant of the third body. For the Sun and the Moon we have
\begin{eqnarray}
\mu_S&=&132712440041.939400 \; \mbox{km$^3/$s$^2$}\; ,  \\
\noalign{\smallskip}
\mu_M&=&4902.800066 \; \mbox{km$^3/$s$^2$}\; ,
\end{eqnarray}
according to the most precise determinations of these paremeters \cite{DE431}.

The last term in the right-hand side of Eq. (\ref{eqmotionv}) arises because the field of the Earth is described in terms of a
series of zonal, tesseral and sectorial harmonics. All these terms must be taken into account because of the polar flattening of the planet and the inhomogeneous distribution of mass inside the planet or the irregularities of the surface. In particular, the distribution of oceans and land masses. Many studies of the perturbations exerted upon the orbits of artificial satellites
throughout the years of the space age, and gravimetry analysis as well, have allowed for the development of an accurate geopotential model. A great achievement was the publication of the NASA GSFC and NIMA joint geopotential model in 1996 
\cite{EGM96} (also known as EGM96). This model is complete up to order $360$ of the harmonic expansion for the potential
in spherical coordinates:
\begin{equation}
\label{UEGM96}
\begin{array}{rcl}
U(r,\theta,\phi)&=&-\displaystyle\frac{\mu_E}{r} \, \displaystyle\sum_{n=2}^N \, \displaystyle\sum_{m=0}^n \, \left( \displaystyle\frac{R}{r} \right)^n P_{n,m}(\cos \theta) \\
\noalign{\smallskip}
& & \left( C_{n,m} \, \cos\left( m \lambda\right)+ S_{n,m} \, \sin\left( m\lambda\right)
\right) \; ,
\end{array}
\end{equation}
where $\theta$ is the polar angle (or colatitude), $\lambda$ is the terrestrial longitude and $R=6378.1363$ km is a reference
radius. Notice that here $P_{n,m}(x)$, $m=0,\ldots,n$ are the associated Legendre functions obtained by application of an extension of Rodrigues' formula:
\begin{equation}
\label{Rodrigues}
P_{n,m}(x)=\displaystyle\frac{(-1)^m}{2^n n !} \, \left(1 - x^2 \right)^{m/2} \, \displaystyle\frac{d^{n+m}}{d x^{n+m}} 
\left( x^2 -1 \right)^n \; .
\end{equation}
Although there are more recent geopotential models, as the one was developed in 2008 (EGM2008) which includes spherical harmonics of degree and order
$2159$, we will see that for our purpose it is enough to take into account the order $N=360$ provided by the EGM96 model
\cite{EGM96}. The latest gravity missions: GRACE, CHAMP and GOCE have
provided additional information of the Earth's gravity field model and the accuracy has been improved (see ICGEM webpage \cite{ICGEM}) but we will show that for our main objective of elucidating the existence of an anomalous acceleration of order
$0.1$ mm$/$s$^2$ acting upon the spacecraft on the vicinity of the Earth, the EGM96 provides sufficient precision and accuracy.

Notice also that in Eq. (\ref{UEGM96}) we are considering only the perturbations to the simple Newtonian
potential $U_0=-\mu_E/r$. In this expansion the zonal, nonzero tesseral and sectorial coefficients of order $n=2$ are, for example:
\begin{eqnarray}
\label{C20}
C_{2,0}&=&-1.08262668 \times 10^{-3} \pm 7.962 \times 10^{-11} \; , \\
\noalign{\smallskip}
C_{2,1}&=&-2.414000000 \times 10^{-10} \pm 1.290 \times 10^{-30} \; , \\
\noalign{\smallskip}
C_{2,2}&=&1.5744603745 \times 10^{-6} \pm 3.468 \times 10^{-11} \; , \\
\noalign{\smallskip}
S_{2,1}&=&1.5431000000044 \times 10^{-9} \pm 1.2909 \times 10^{-30} \; , \\
\noalign{\smallskip}
\label{S22}
S_{2,2}&=&-9.038038 \times 10^{-7} \pm 3.5084 \times 10^{-11} \; .
\end{eqnarray}
We must also remember that among the coefficients as they appear in the model given by Eq. (\ref{UEGM96}) and the tabulated
ones by Lemoine et al. \cite{EGM96} there is a conversion factor:
\begin{equation}
\label{CSconv}
C_{n,m}=\kappa(m) (2 n +1) \displaystyle\frac{(n-m) !}{(n+m) !} \bar{C}_{n,m} \; ,
\end{equation}
where $\bar{C}_{n,m}$ are the tabulated values and $\kappa(m)=1$ for $m=0$ and $\kappa(m)=2$ for $m \neq 0$. A similar
conversion is necessary for $S_{n,m}$.
From Eq. (\ref{UEGM96}) we can find now the components of the perturbing force in spherical coordinates:
\begin{eqnarray}
\label{Fgeor}
{\mathcal F}_r&=&-\displaystyle\frac{\partial U}{\partial r} \\
\noalign{\smallskip}
&=&-\displaystyle\frac{\mu_E}{r^2} \, \displaystyle\sum_{n=2}^N\,
\displaystyle\sum_{m=0}^n \, (n+1)\left( \displaystyle\frac{R}{r} \right)^n \, P_{n,m}(\cos \theta) \\
\noalign{\smallskip}
& &\left\{ C_{n,m} \cos \left( m
\lambda \right)+S_{n,m} \sin\left( m \lambda \right) \right\} \; ,\\
\noalign{\smallskip}
\label{Fgeot}
{\mathcal F}_\theta&=& -\displaystyle\frac{1}{r} \, \displaystyle\frac{\partial U}{\partial \theta} \\
\noalign{\smallskip}
&=&-\displaystyle\frac{\mu_E}{r^2}\, \displaystyle\sum_{n=2}^N  \, \displaystyle\sum_{m=0}^n \, \left(\displaystyle\frac{R}{r} \right)^n\, P^{'}_{n,m}(\cos \theta) \, \sin\theta \\
\noalign{\smallskip}
& &\left\{ C_{n,m} \cos \left( m
\lambda \right)+S_{n,m} \sin\left( m \lambda \right) \right\} \; , \\
\noalign{\smallskip}
\label{Fgeol}
{\mathcal F}_\lambda &=& -\displaystyle\frac{1}{r \sin\theta} \, \displaystyle\frac{\partial U}{\partial \lambda} \\
\noalign{\smallskip}
&=&\displaystyle\frac{\mu_E}{r^2 \sin \theta} \, \displaystyle\sum_{n=2}^N \, \displaystyle\sum_{m=1}^n \, m 
 \left(\displaystyle\frac{R}{r} \right)^n P_{n,m}(\cos \theta)\\
\noalign{\smallskip} 
& &\left\{ - C_{n,m} \sin \left( m
\lambda \right)+S_{n,m} \cos\left( m \lambda \right) \right\} \; .
\end{eqnarray}
The total perturbing force vector is then given by $\bm{\mathcal{F}}_{\mbox{geo}}=
\mathcal{F}_r \, \hat{r}+\mathcal{F}_\theta \, \hat{\theta}+\mathcal{F}_\lambda \, \hat{\lambda}$, where the unit vector
$\hat{\lambda}$ points to the west and, consequently, opposite to the Earth's rotation. In the following section, we will
discuss the sources of error of the integration performed with Eqs. (\ref{eqmotionr})-(\ref{Ftid}) and the
components of the geopotential perturbation in Eqs. (\ref{Fgeor})-(\ref{Fgeol}).
\subsection{Error analysis}
Our objective is to disclose any anomalous component in the force acting upon a spacecraft which flybys the Earth once the
sources of classical perturbations have been taken into account. The magnitude of the expected anomalous acceleration can be
estimated from some models that have been proposed before to study the anomaly \cite{Acedo2015,Acedo2017three}. In these models
an acceleration of magnitude:
\begin{equation}
\label{aestimate}
\delta a = \displaystyle\frac{ \mu_E}{R_E^2} \, \displaystyle\frac{\Omega R_E}{c}  \simeq 1.52 \times 10^{-8} \; \mbox{km$/$s$^2$}\; ,
\end{equation}
is found as the source of the anomaly, where $R_E =6371$ km  is the average radius of the Earth and $\Omega R_E/c \simeq 1.5495 \times 10^{-6}$ s$^{-1}$ is the ratio of the linear velocity of a point at the Earth's equator, as a consequence of the Earth's rotation around its axis, and the speed of light. Therefore, we should show that all the sources of error are, in order of magnitude, negligible or, at least, small in comparison with the expected value of the anomalous acceleration we suppose to be acting upon the spacecraft in the vicinity of the perigee.
\subsubsection{Atmospheric friction}
First, we will analyze the kinematic effect of atmospheric friction on the proximity of the perigee for a typical flyby. We will
show that for altitudes over $500$ km the deceleration is negligible because the density of the thermosphere at those altitudes
is very low. In particular, for the NEAR flyby the impact of this effect on the final outgoing velocity has been evaluated in the range of a hundredth of mm$/$s \cite{Acedo2017one}.

The drag force due to the atmospheric friction is estimated by the usual expression \cite{KingHele}:
\begin{equation}
\label{drag}
{\bf D}=-\displaystyle\frac{1}{2} \, \rho v^2 A C_d \bm{\hat{v}}\; ,
\end{equation}

where $\rho$ is the atmospheric density, $v$ is the spacecraft velocity, $A$ is the cross-sectional area perpendicular to the direction of motion and $C_d$ is the drag coefficient (which for satellites and various spacecraft geometries \cite{Moe} is roughly $C_d \gtrsim 2$). The required parameters for NEAR at perigee (attained on January 23rd, 1998 at $7$:$24$ UTC) are: the mass, $m=730$ kg, velocity modulus, $V_p=12.739$ km$/$s, effective area, $A C_d\simeq 2*1.5*2.75$ m$^2$, altitude at perigee, $h=539$ km, atmospheric density, $\rho=1.133 \times
10^{-13}$ kg$/$m$^3$. Inserting these values into Eq. (\ref{drag}) yields:
\begin{equation}
a_{\mbox{Friction}} = 1.0244 \times 10^{-10} \; \mbox{km$/$s$^2$}\; ,
\end{equation}
which it is two orders of magnitude smaller than the expected anomalous acceleration in Eq. (\ref{aestimate}). On the other hand, there is, at least, one flyby in the series we are considering in which the effect of atmospheric friction cannot be neglected.
For the second Galileo flyby of the Earth in December 8th, 1992 the minimum altitude was $303$ km and, in this case, a total
decrease of the final asymptotic velocity of $-4.6$ mm$/$s (from the observed $-8$ mm$/$s) is attributed only to friction
\cite{Anderson2008,Acedo2017one}. This effect is to be carefully incorporated into the orbital model for this flyby in order
to disclose any remaining anomalous acceleration.
\subsubsection{Tidal forces exerted by other planets}
In our orbital model for the flyby trajectory around the perigee, we are going to ignore tidal forces arising from other celestial
bodies apart from the Sun and the Moon. However, we must check that these are sufficiently small to be safely excluded from
the analysis. The third body in importance for tidal perturbations of Earth flybys is Jupiter. The coordinates of NEAR at perigee
are \cite{Horizons}:
\begin{eqnarray}
\label{NEARcoor}
X_{\mbox{NEAR}}&=&1042.0129 \; \mbox{km} \; ,\\ 
\noalign{\smallskip}
Y_{\mbox{NEAR}}&=&-3750.0865 \; \mbox{km} \; ,\\
\noalign{\smallskip}
Z_{\mbox{NEAR}}&=&5710.3327 \; \mbox{km}\; .
\end{eqnarray}
Where we are using the ecliptic reference frame with origin at the center of the Earth.
At this instant, the position of Jupiter was:
\begin{eqnarray}
\label{JUPcoor}
X_{\mbox{Jupiter}}&=&4.95651 \; \mbox{AU} \; ,\\ 
\noalign{\smallskip}
Y_{\mbox{Jupiter}}&=&-3.19481 \; \mbox{AU} \; ,\\
\noalign{\smallskip}
Z_{\mbox{Jupiter}}&=&-0.08915 \; \mbox{AU}\; .
\end{eqnarray}
Here, $AU=149597870.7$ km is the astronomical unit. Finally, the mass term for Jupiter is $\mu_J=126712764.8$ km$^3/$s$^2$.
With these data, we can calculate the tidal acceleration at perigee from Eq. (\ref{Ftid}) yielding $a_{\mbox{tidal}} = 1.6278 \times 10^{-10}$ km$/$s$^2$, i.e., two orders of magnitude below the relevant acceleration magnitude we are looking for in the 
flyby data.
\subsubsection{Mismodelling of the zonal, tesseral and sectorial harmonics and geocentric latitude and longitude}
In the EGM96 and EGM2008 models error bars for the uncertainty of the zonal, tesseral and sectorial coefficients of the geopotential
model are provided. In Eqs. (\ref{C20})-(\ref{S22}), some of these error bars are given. If we assume that these errors add
up in the same direction, we can estimate the maximum error in the calculation of the perturbing forces corresponding to the
geopotential model from Eqs. (\ref{Fgeor})-(\ref{Fgeol}) and this yields the following result
for the expected maximum error at the perigee of the NEAR flyby:
\begin{equation}
\left\vert \delta \bm{\mathbf F}_{EGM96} \right\vert = 4.399 \times 10^{-11} \; \mbox{km$/$s$^2$} \; ,
\end{equation}
which it is clearly below the threshold of the required precision. 
Another issue with the application of the geopotential model is that we must know the geocentric colatitude, $\theta$, and the
longitude, $\lambda$, in order to evaluate the perturbing forces in Eqs. (\ref{Fgeor})-(\ref{Fgeol}). If we know the orientation
of the Earth's rotation axis in the ecliptic frame, $\bm{\hat{k}}$, we can calculate the cosine of the spacecraft colatitude by projecting the
its position vector as follows:
\begin{equation}
\label{colatitude}
\cos \theta=\displaystyle\frac{{\bf r} (t) \cdot \khat}{\left \vert{\bf r}(t) \right \vert} \; .
\end{equation}
We can accurately compute the obliquity of the ecliptic by a polynomial \cite{Almanac}:
\begin{equation}
\label{oblq}
\begin{array}{rcl}
\chi&=& 23^\circ 26^{'} 21.406^{''}-46.836769^{''} T \\
\noalign{\smallskip}
&-&0.0001831^{''} T^2+0.00200340^{''} T^3 \\
\noalign{\smallskip}
&-&5.76^{''}\times 10^{-7} T^4-4.34^{''} \times  10^{-8} T^5\; ,
\end{array}
\end{equation}
where $T$ is the time from the epoch J2000.0 in Julian centuries. For example, we have $T=-708/365.25$ for the day of the NEAR
flyby and this yields $\chi=23^\circ 26^{'} 22.3139^{''}$. The relation among the equatorial unit vectors, $\ihat$, $\jhat$ and
$\khat$, and the ecliptic ones, $\bf{\hat{e}}_i$, $i=1$,$2$,$3$:
\begin{eqnarray}
\ihat&=& \bf{\hat{e}}_1 \; , \; \jhat= \bf{\hat{e}}_2\; ,   \\
\noalign{\smallskip}
\khat&=& \sin \chi \, \bf{\hat{e}}_2+\cos \chi \, \bf{\hat{e}}_3 \; .
\end{eqnarray}
The present models for the evolution of the axial tilt of the Earth provide and accuracy of $10^{-4}$ seconds of arc. Anyway,  we notice that, even for an error of one second of arc, a sensitivity analysis using Eqs. (\ref{Fgeor})-(\ref{Fgeol}) and the 
computation of the cosine of the colatitude in Eq. (\ref{colatitude}) shows that the uncertainty in the acceleration imparted
by the perturbing terms of the geopotential is changed only  by $5.50\times 10^{-10}$ km$/$s$^2$ for the NEAR flyby and this is 
only a one per cent of the anomaly we are expecting to find. We should also emphasize that the transformation
of coordinates would ideally follow the IERS2010 convention \cite{IERS2010} including the effects of precession, nutation, polar motion, etc but such an accuracy is not required in our model for the same reasons discussed above.

The terrestrial longitude of the vertical of the spacecraft position, $\lambda(t)$, is related to the right ascension, $\alpha(t)$, by:
\begin{equation}
\label{lambda}
\lambda(t) = \lambda_0 -\Omega t+\alpha(t)-\alpha_0 \; ,
\end{equation}
$\Omega$ being the angular velocity of the Earth's rotation around its axis and $\alpha(t)$ the right ascension of the spacecraft.
The parameters $\lambda_0$ and $\alpha_0$ are the geocentric longitude and right ascension of the spacecraft at $t=0$. The last
two parameters are connected by the expression:
\begin{equation}
\label{lambdazero}
\lambda_0=\alpha_0-\mbox{LST(Greenwich)} \; ,
\end{equation}
where LST(Greenwich) is the local solar time at Greenwich, U. K., i. e., the right ascension of a point at the Greenwich meridian. Notice that the right ascension and the geocentric longitude are measured counterclockwise as seen from the North Pole.
For the NEAR flyby occurred on January 23rd, 1998 at precisely $7$:$24$ UTC we have that the local solar time of 
Greenwich was $\mbox{LST(Greenwich)}=15$ hours, $33$ minutes and $44$ seconds and for the right ascension of the spacecraft, at the
closest instant to perigee, we can find in the Horizons ephemeris program \cite{Horizons} that it was $18$ hours, $30$ minutes and $28.23$ seconds. This gives, according to Eq. (\ref{lambdazero}), that $\lambda_0=44.184291^\circ$ in sexagesimal degrees for the vertical of the NEAR spacecraft at perigee. The uncertainty in $\lambda_0$ is, then, at most one second of time, i.e., approximately $\Delta \lambda_0=0.0042^\circ$ in sexagesimal degrees. Consequently, the error bars in the geopotential perturbing force modulus evaluated from
Eqs. (\ref{Fgeor})-(\ref{Fgeol}) is bounded by $\left\vert \delta \bm{\mathcal F}_{\mbox{EGM96}} (\Delta \lambda_0)\right\vert <
1.69 \times 10^{-11}$ km$/$s$^2$ and it is also sufficiently small to be of no serious concern in our orbital analysis
for the NEAR and other flybys.

Another important issue is the temporal variation of the Earth's oblateness as a consequence of large-scale mass
transports \cite{Chao2006}. The temporal variation of $J_2$ is of particular interest in this context. The variation of this
coefficient is correlated with mass transport in the atmosphere, oceans and land hydrology and its amplitude is given, approximately, by $\Delta J_2 \simeq 3 \times 10^{-10}$. A decreasing trend at a rate of $-2.8 \times 10^{-11}$ yr$^{-1}$ has also been detected since the seventies of the past century but these variations are still very small to contribute significantly to the spacecraft acceleration as they imply only errors $\sim 10^{-4}$ mm$/$s$^{-2}$ in our simulation of the anomalous accelerations. This is, obviously, too small to explain away the flyby anomaly as we will see in the next section.

\subsubsection{Error control in the numerical methods}

To solve the equations of motion we use a Picard iterative method \cite{Picard} in which each iteration is integrated numerically.
So for the $n$-th iteration we have:
\begin{equation}
\label{EqPic}
\displaystyle\frac{d^2 {\bf r}}{d t^2}+\mu_E \,\displaystyle\frac{{\bf r}}{r^3}=\displaystyle\sum_i \, \bm{\mathcal F}_i\left ({\bf r}_{n-1}\right) \; ,
\end{equation}
where the right-hand side correspond to the sum over all perturbing forces evaluated at the already known positions, ${\bf r}_{n-1}(t)$ for the previous iteration $n-1$. The differential equation in Eq. (\ref{EqPic}) is then solved numerically to
obtain the $n$-th order approximation, ${\bf r}_n(t)$ and the process is repeated until we achieve the desired accuracy. In our case we start with ${\bf r}_0(t)$ as the keplerian solution in the absence of any perturbing forces. After iterating two times, we have found that the errors are so small that no more iterations are necessary. For example, $6$ minutes after the perigee of the NEAR flyby we find that:
\begin{eqnarray}
\left\vert \left\vert \bm{\mathcal F}({\bf r}_1)\right\vert -\left\vert \bm{\mathcal F}({\bf r}_0)\right\vert\right\vert &\simeq&
4.77  \times 10^{-9} \; \mbox{km$/$s$^2$} \; ,\\
\noalign{\smallskip}
\left\vert {\bf r}_1 \right\vert -\left\vert {\bf r}_{\mbox{obs}}\right\vert &=& 10.1234 \; \mbox{m} \; ,
\end{eqnarray}
for the difference (in modulus) among the perturbing forces at the keplerian positions, ${\bf r}_0$ and the corrected ones, ${\bf r}_1$. We also find that the difference of the corrected trajectory with the observations, ${\bf r}_{\mbox{obs}}$ is only 
of the order of $10$ meters. In the second iteration we have that the difference among the perturbing forces evaluated
at positions ${\bf r}_1$ and ${\bf r}_2$ are greatly reduced:
\begin{equation}
\left\vert \left\vert \bm{\mathcal F}({\bf r}_2)\right\vert -\left\vert \bm{\mathcal F}({\bf r}_1)\right\vert\right\vert \simeq
9.02  \times 10^{-14} \; \mbox{km$/$s$^2$} \; ,
\end{equation}
and this is already six orders of magnitude smaller than the magnitude of the anomalous acceleration we are investigating in this paper. Concerning the difference in positions we have $\vert \vert {\bf r}_2 \vert -\vert {\bf r}_1 \vert\vert=2.52$ cm, which it is sufficiently small to justify the use of only two iterations of the Picard method for the evaluation of the flyby orbit in the
time interval of our interest.

Integration of the equations of motion was carried out in Mathematica \cite{Math11} with standard routines and double precision.

\section{Results}
\label{results}
In this section we will show that an unexplained residual discrepancy remains among the position provided by the Deep Space Network telemetry data and the positions and accelerations predicted by orbital models, after considering all the classical effects:
\begin{itemize}
\item Atmospheric friction \cite{Acedo2017one}.
\item Tidal forces by the Sun, the Moon and the planets.
\item Zonal, tesseral and sectorial contributions from the geopotential models \cite{AcedoMNRAS}.
\item Solid and ocean tides.
\item Corrections provided by General Relativity which, as a rule of thumb \cite{WillPN}, are ${\mathcal O}(v^2/c^2 g)\simeq 10^{-9} g\simeq 10^{-11} \mbox{ km$/$s$^2$}$ for a typical velocity at perigee of $10$ km$/$s.
\item Other minor effects already studied and dismissed by other authors (spacecraft charge \cite{LPDSolarSystem}, Lense-Thirring or gravitomagnetic field of the Earth \cite{IorioSRE2009,Hackmann}, solar wind and anisotropic thermal emission\cite{Rievers2011}).
\end{itemize}
The IERS 2010 conventions \cite{IERS2010} also recommend to take into account effects such as the solid Earth and 
ocean pole tides. These are a consequence of the ocean response to the small perturbations to the Earth's rotation axis that primarily occur with a period of $433$ days (the so-called Chandler's wobble) \cite{Desai2002}. However, these correspond only to
a change in the altitude of the tide only of the order of one centimente and the effect of such changes are totally negligible for the analysis of the flyby anomaly (See \cite{AcedoMNRAS,LPDSolarSystem}).

Solar radiation pressure, Earth's infrared radiation pressure and albedo modelling are other radiative effects that could play a role in the study of high-precision orbital dynamics. In particular, it would be interesting to take into account its variation
depending upon the Sun's location in the sky as the spacecraft performs its flyby. Nevertheless, these effects are expected to be
very small in comparison with the magnitude of the anomalous acceleration responsible for the flyby anomaly ($\simeq 0.1$ 
mm$/$s$^2$). Estimations of L\"ammerzahl et al. \cite{LPDSolarSystem} are $\simeq 2.4 \times 10^{-6}$ mm$/$s$^2$ for both the 
Earth albedo and Solar wind effect. Consequently, we can safely ignore these minor effects in our orbital model for the flyby
anomaly.

Another radiative effect is the one suggested and studied by I. V. Yarkovsky in the late XIXth century and the beginning of the XXth century. According to this author, we should find that small asteroids or spacecraft would suffer a speed-up or speed-down of their rotational velocities as a result of the asymmetric emission of infrared radiation from their surfaces as they are heated by the Sun. This effect is expected to be important for large time-scales of thousands or millions of years \cite{Rubincam2000} but we can ignore it on a short-time process such as a flyby whose duration is measured in hours.

The induction of currects as the spacecraft crosses the magnetic field of the Earth is another small effect to be considered. For a surface of $S \simeq 10$ m$^2$, we can estimate the maximum magnetic flux as $\varphi \simeq B_{\mbox{max}} S$, with $B_{\mbox{max}} = 65 \times 10^{-5}$ Teslas at the surface of the Earth. A bound on the maximum current induced can then be obtained by assuming that the flux changes from this maximum to zero during the time that the spacecraft spends in the vicinity of the Earth's surface (approximately, $T=1$ hour). For a minimum resistance of $R=0.001$ ohms, corresponding to a thick copper wire, we get $I_{\mbox{max}} = \varphi_{\mbox{max}} / T R \simeq 1.80 \times 10^{-4}$ A and a magnetic moment $m = 2 I_{\mbox{max}} S \simeq 0.036$ A m$^2$. According to L\"ammerzahl et al.  \cite{LPDSolarSystem}, the steepness of the magnetic field close to the Earth is not larger than $\vert \Delta B / \Delta x \vert \simeq 2\times 10^{-7}$ gauss/m. The force for the spacecraft with magnetic moment $\mathbf{m}$ would be $\mathbf{F}= \mathbf{\nabla} \left( \mathbf{m} \cdot \mathbf{B}\right)$ and, in magnitude, is not larger than $\vert \mathbf{F}\vert < 7.22 \times 10^{-13}$ N. For a mass of $10^3$ Kg this leads to an 
acceleration of the order of $10^{-13}$ mm$/$s$^2$ which is far too small in comparison with the expected anomalous acceleration in the flyby anomaly.

\begin{figure}
\includegraphics[width=\columnwidth]{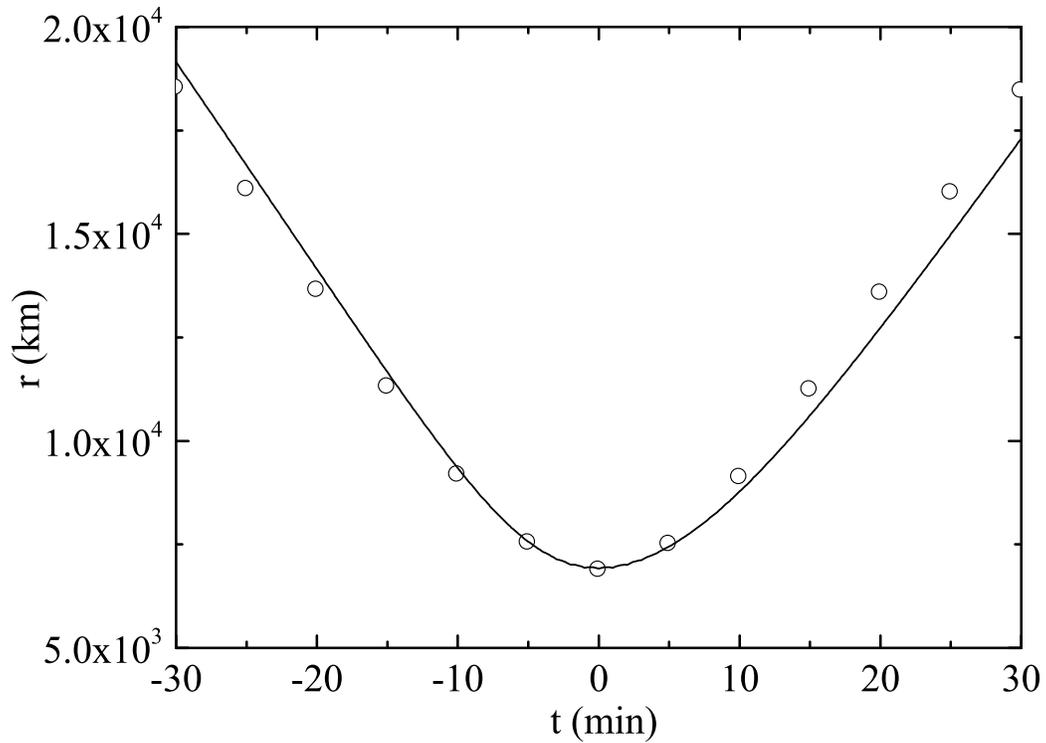}
\caption{The modulus of the position vector of the spacecraft and the prediction from the Keplerian model of the orbit in km vs time in minutes ($t=0$ corresponds to the point in the data closer to the surface of the Earth). The real orbit is plotted as a solid lines and the keplerian approximation as circles. The difference among them has been enlarged by a factor $10^3$ to enhance  the discrepancy which it is shown below on a separate figure.}
\label{fig:1}       
\end{figure}

\begin{figure}
\includegraphics[width=\columnwidth]{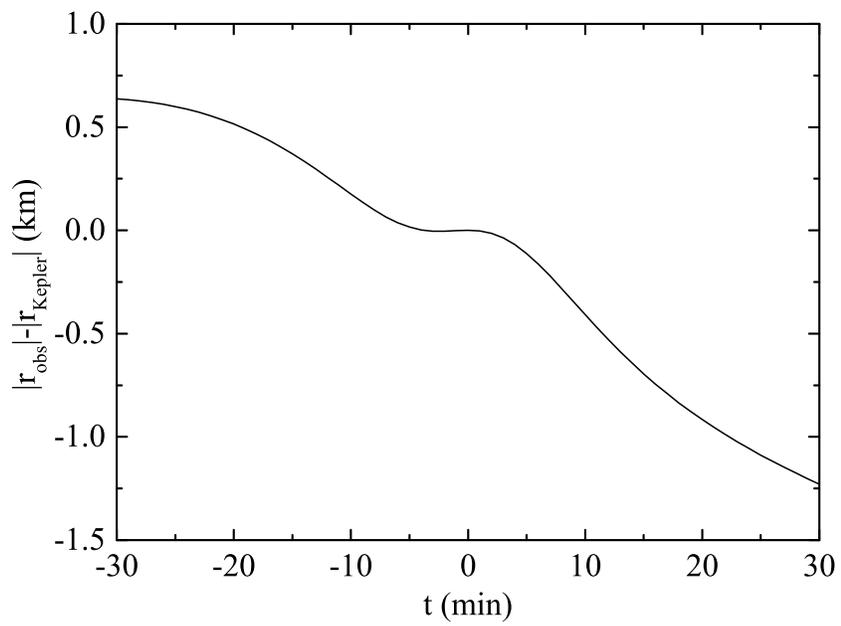}
\caption{Difference among the modulus of the position vector of the spacecraft and the prediction from the Keplerian model of the orbit in km vs time in minutes.}
\label{fig:2}       
\end{figure}

The residual discrepancy (after taking into account all these effects or dismissing them because of its small magnitude) allows us to estimate the anomalous acceleration acting upon the spacecraft as a function of time around its perigee. To do so, we start by calculating the first iteration of the equation of motion, i. e., we obtain the keplerian orbit.
In Fig.\ \ref{fig:1}, we have plotted the predictions of this keplerian orbit for the NEAR flyby in comparison with the real data obtained from the NASA database \cite{Horizons}. Apparently, the agreement is very good but this is consequence of the large distance scale used in the diagram. If we plot the difference among the modulus of the real position vector and the keplerian model we obtain the results in Fig. \ref{fig:2} where we see that difference up to one km builds up during a period of $30$ minutes after the perigee. 

\begin{figure}
\includegraphics[width=\columnwidth]{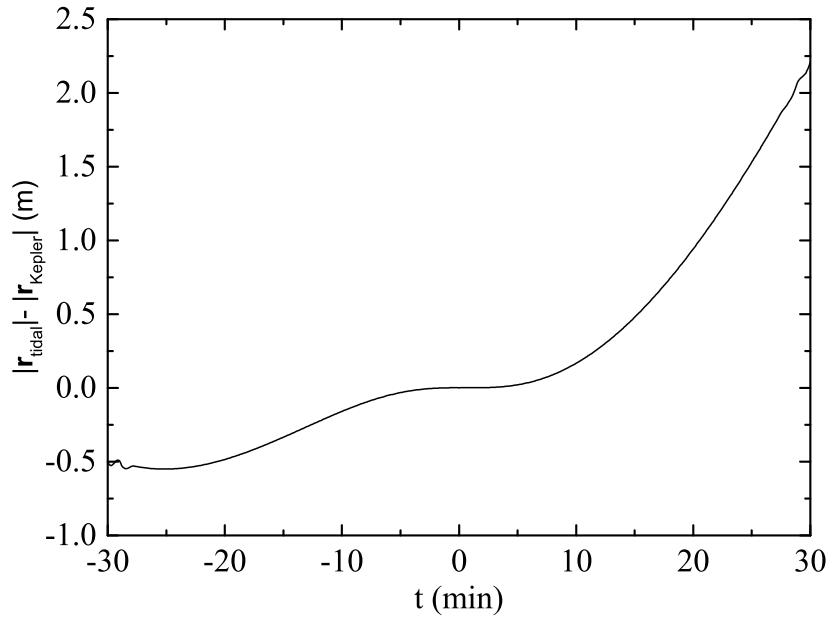}
\caption{Difference among the modulus of the position vector of the spacecraft in the model with tidal forces and the prediction from the Keplerian model of the orbit in meters vs time in minutes. Notice that the effect of incorporating the tidal forces from the Sun and the Moon is three orders of magnitude smaller than the discrepancies of the real data with the Keplerian orbit.}
\label{fig:3}       
\end{figure}

\begin{figure}
\includegraphics[width=\columnwidth]{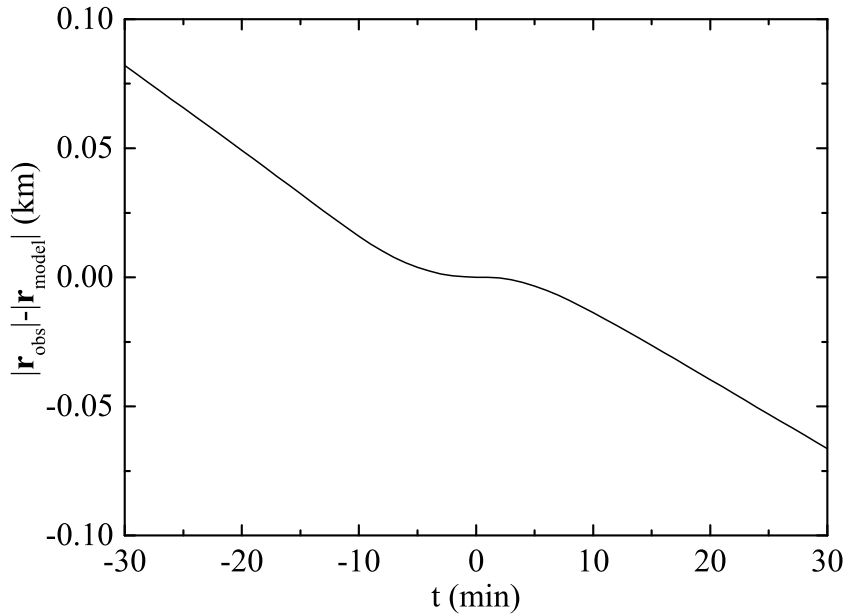}
\caption{Difference among the modulus of the position vector of the spacecraft for the real data and the prediction from the model of the orbit with the tidal and geopotential perturbations (in km) vs time in minutes. This discrepancy corresponds to the anomaly we are looking for in our analysis.}
\label{fig:4}       
\end{figure}

To reduce this discrepancy we have incorporated, in a first step, the tidal forces of the Sun of the Moon as given by Eq.\ (\ref{Ftid}) and the result is shown in Fig. \ref{fig:3} where we have plotted the difference among the vector modulus for the
model with tidal forces and the ideal keplerian orbit. It is clear that this contribution is completely insufficient (a few
meters in comparison with some kilometers) to explain the data, apart from being of the wrong sign. This shows that the contributions of the perturbation arising from the terms in the geopotential model should be further more important.
Finally, we have incorporated also the geopotential terms in Eqs. (\ref{Fgeor})-(\ref{Fgeol}). By computing the difference of the 
real distance to the center of the Earth and the prediction of the model we obtain the results in Fig.\ (\ref{fig:4}). We find that, after the perigee, the NEAR spacecraft was closer to the Earth than we should expect according to our orbital model. The difference being $50$ m after, approximately, $25$ minutes. This discrepancy we have found in the NEAR flyby is also found in other flybys and points towards an anomaly in the reconstruction of the orbit that cannot be eliminated by conventional effects in perturbation theory.
Now that we have analyzed all the sources of perturbation and errors in the evaluation of the trajectory of the NEAR spacecraft in the vicinity of its perigee. So, we can estimate the residual acceleration giving rise to the discrepancies in the coordinate's data. If we denote by $\delta {\bf r}$ the difference among the real and the predicted position, as plotted in Fig. \ref{fig:4},
we have that a good approximation to the extra acceleration corresponding to this discrepancy is given by a fourth-order finite difference method \cite{Fornberg} as follows:
\begin{equation}
\label{deltaa}
\begin{array}{rcl}
\delta{\bf a}&=&\displaystyle\frac{1}{h^2} \,\left\{ -\displaystyle\frac{1}{12}\left( \delta {\bf r}(t - 2 h)+
\delta {\bf r}(t+2 h)\right)\right. \\
\noalign{\smallskip}
&+&\left. \displaystyle\frac{4}{3}\left( \delta {\bf r}(t-h)+\delta {\bf r}(t+h) \right)-
\displaystyle\frac{5}{2} \delta {\bf r}(t)\right\} \\
\noalign{\smallskip}
&-&\displaystyle\frac{1}{90} \displaystyle\frac{d^6 \delta {\bf r}}{d t^6} \, h^4+ {\mathcal O}\left(h^5\right) \; ,
\end{array}
\end{equation}
where $h$ is the timestep which we will take as $h=1$ min to conform to the real time interval used in the spacecraft tracking. Similarly, we can use a second-order central finite difference method to evaluate the sixth derivative in Eq. (\ref{deltaa}) to give an estimation of the numerical error in the numerical method:
\begin{equation}
\label{deltaerr}
\begin{array}{rcl}
\displaystyle\frac{d^6 \delta {\bf r}}{d t^6}&=& \displaystyle\frac{1}{h^6} \left( \delta {\bf r}(t-3 h)+ \delta {\bf r}(t+3 h) 
\right. \\
\noalign{\smallskip}
&-&6 \delta {\bf r}(t-2 h)-6 \delta {\bf r}(t + 2 h)+ 15 \delta {\bf r} (t- h) \\
\noalign{\smallskip}
&+&\left. 15 \delta {\bf r}(t+ h)- 20 \delta {\bf r} (t) \right)+{\mathcal O}\left( h^2 \right)\; ,
\end{array}
\end{equation}
From the discrepancies in the position, with respect to the orbital model, we can calculate the corresponding acceleration vector
and its numerical error from Eqs. (\ref{deltaa})-(\ref{deltaerr}). 
\begin{figure}
\includegraphics[width=\columnwidth]{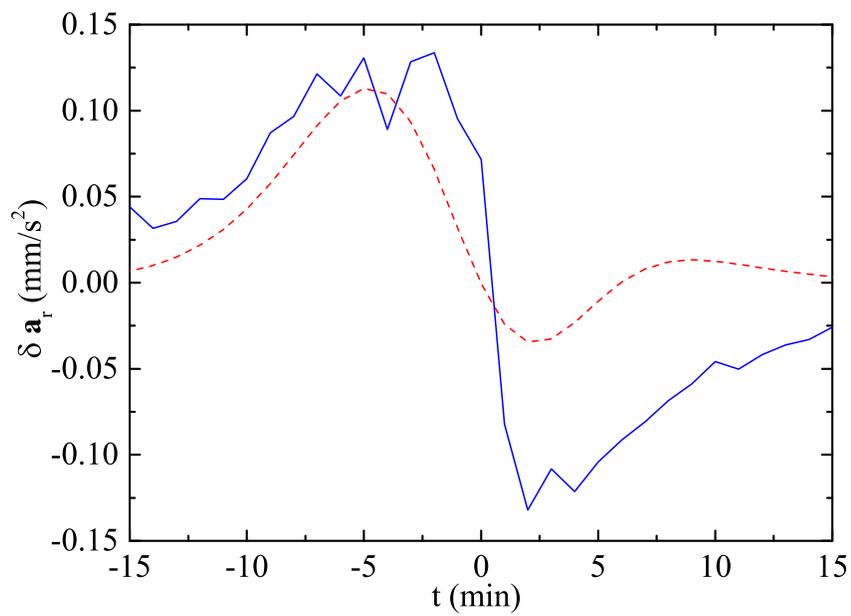}
\caption{Radial component of the anomalous acceleration for the NEAR flyby in mm$/$s$^2$ vs time in minutes where $t=0$ corresponds to the perigee. The solid line is the result of the numerical analysis of the orbital model and the dotted line is
a possible fitting as discussed in the main text.}
\label{fig:5}       
\end{figure}
\begin{figure}
\includegraphics[width=\columnwidth]{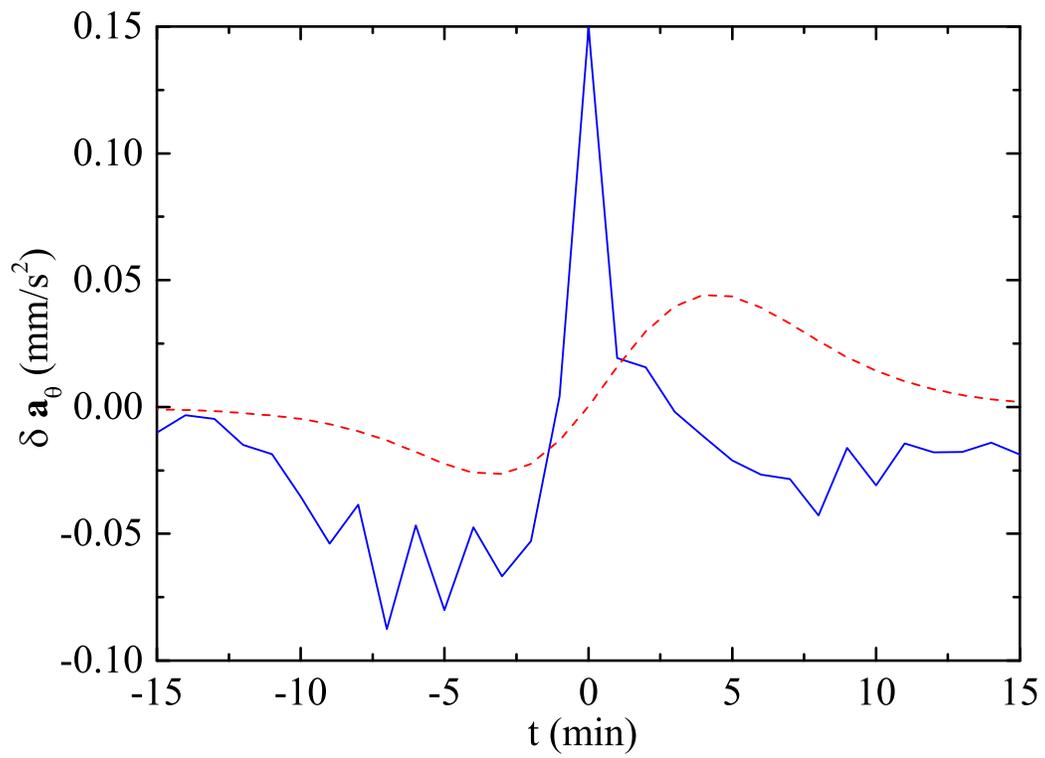}
\caption{The same as Fig. \protect\ref{fig:5} but for the polar component.}
\label{fig:6}       
\end{figure}
\begin{figure}
\includegraphics[width=\columnwidth]{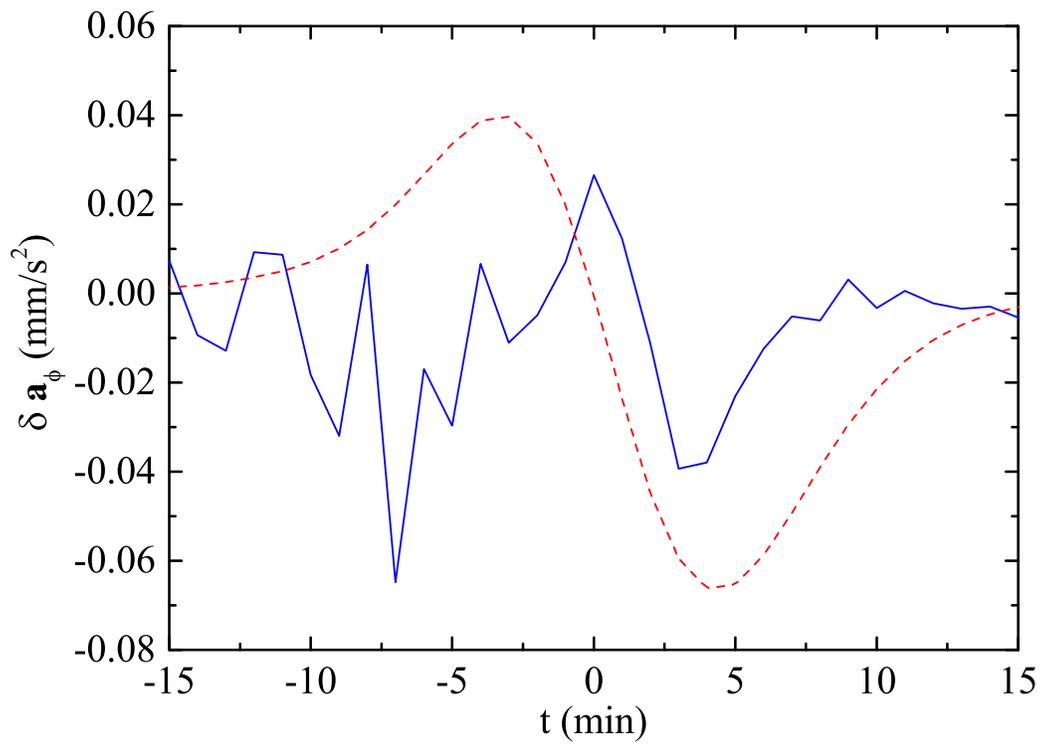}
\caption{The same as Fig. \protect\ref{fig:5} but for the azimuthal component of the anomalous acceleration.}
\label{fig:7}       
\end{figure}
\begin{figure}
\includegraphics[width=\columnwidth]{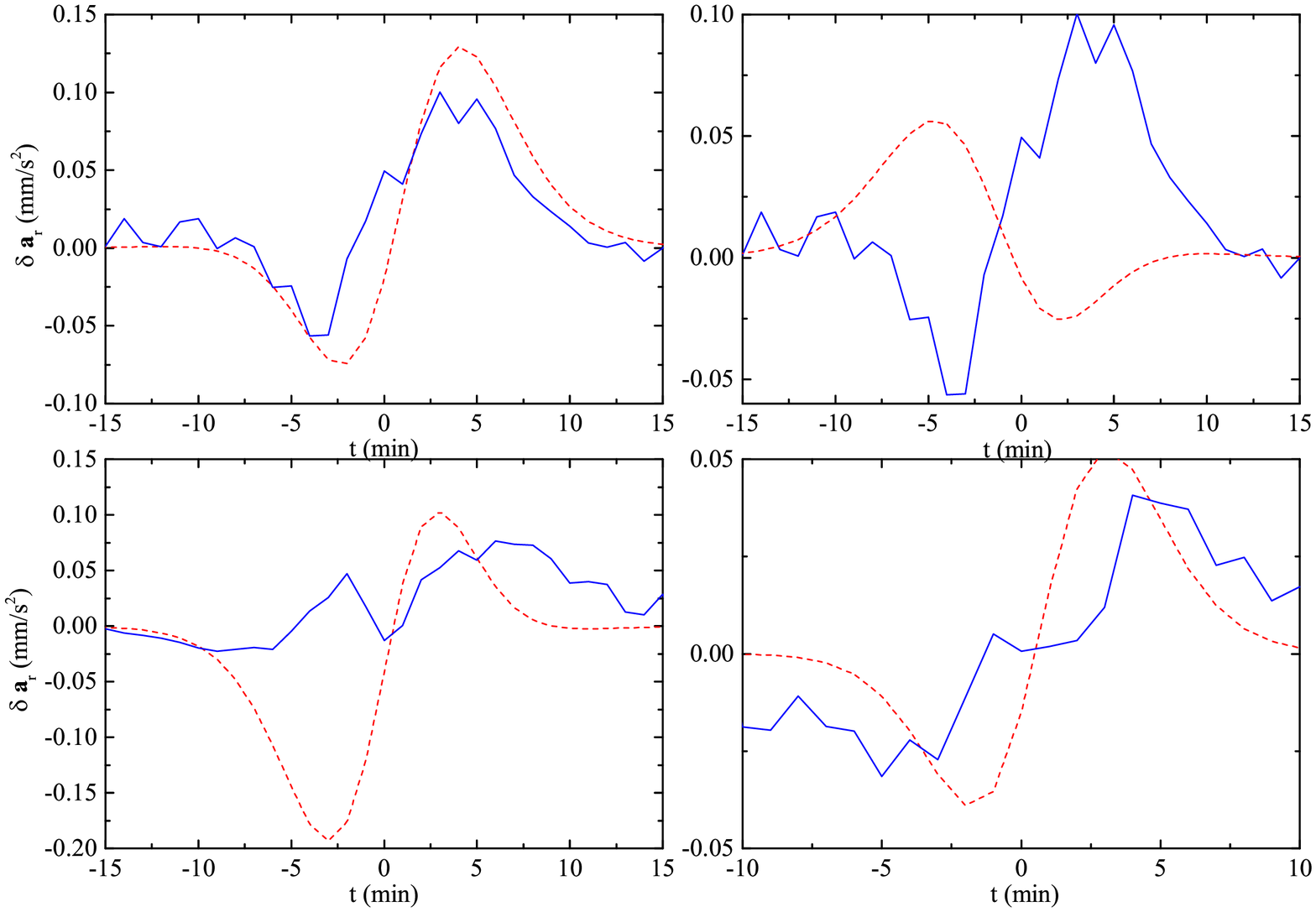}
\caption{Radial component of the anomalous acceleration as derived from our orbital model for the (from left to right and from top to bottom): Galileo II, Galileo I, Juno and Cassini flybys vs time from the respective perigee in minutes. Solid lines are
the numerical results and dotted lines is the prediction of a tentative model discussed in the main text.}
\label{fig:8}       
\end{figure}
\begin{figure}
\includegraphics[width=\columnwidth]{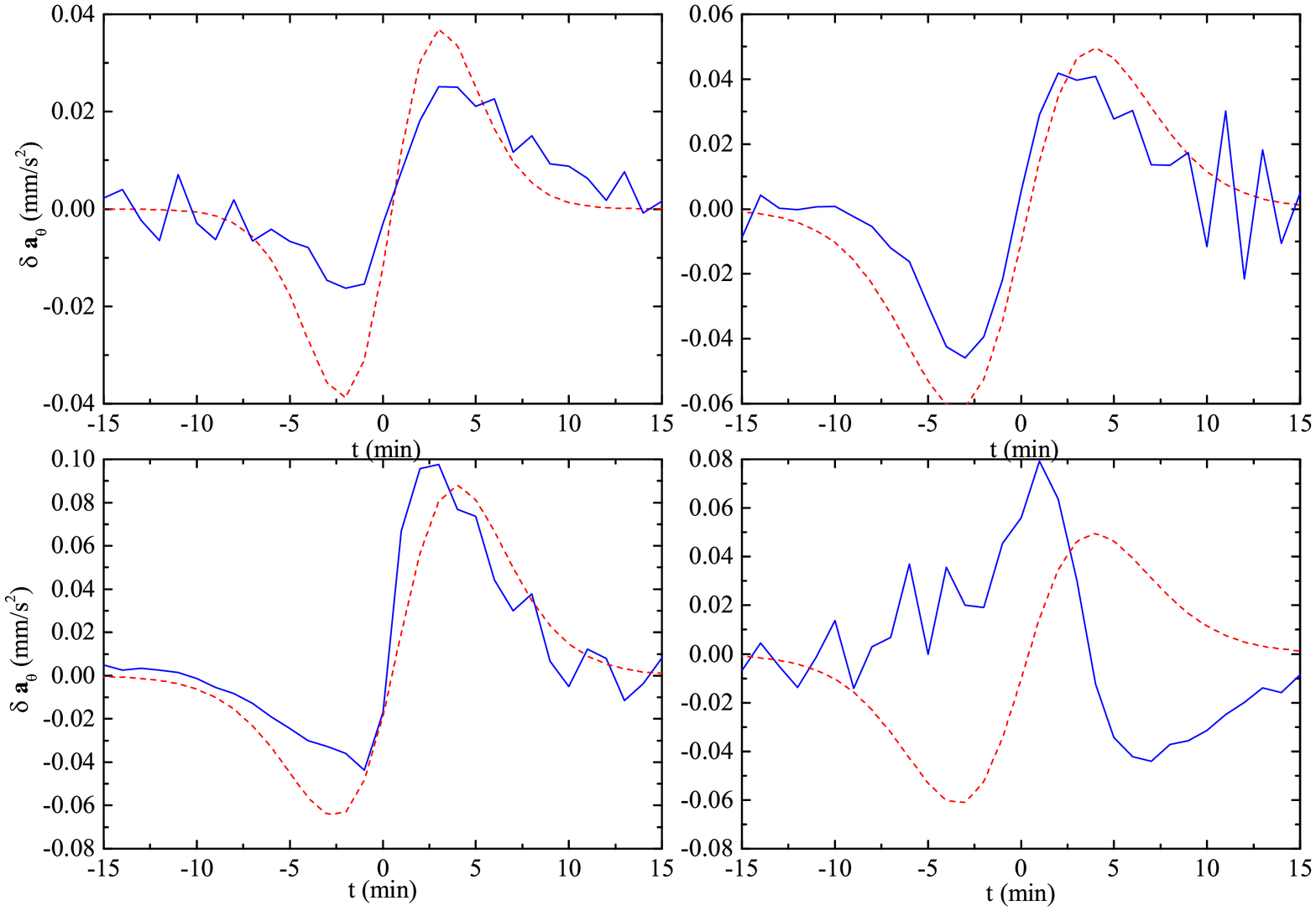}
\caption{Polar component of the anomalous acceleration as derived from our orbital model for the (from left to right and from top to bottom): Cassini, Galileo I, Juno and Galileo II flybys vs time from the respective perigee in minutes. Solid lines are
the numerical results and dotted lines is the prediction of a tentative model discussed in the main text.}
\label{fig:9}       
\end{figure}
\begin{figure}
\includegraphics[width=\columnwidth]{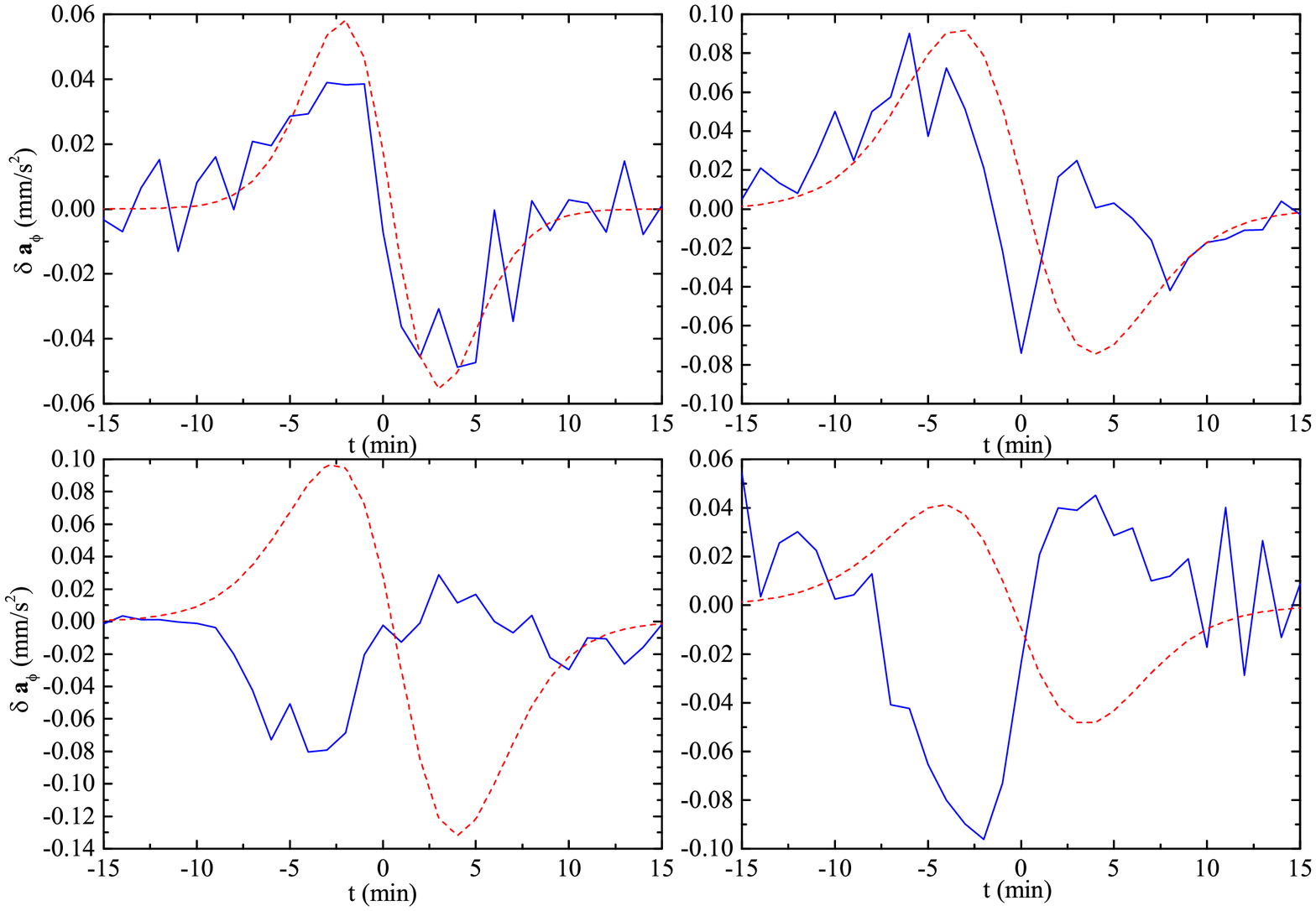}
\caption{The same as Figs. \ref{fig:8} and \ref{fig:9} but for the azimuthal component on the Cassini, Galileo II, Juno and Galileo I flybys.}
\label{fig:10}       
\end{figure}
To keep the precision of the anomalous acceleration as high as possible, it is convenient to choose the time step, $h$, as one minute in Eq. (\ref{deltaa}). In doing so, we obtain the results plotted in Figs. \ref{fig:5}-\ref{fig:7} for the radial, polar
and azimuthal components during the NEAR flyby. In the first place, we notice that the magnitude of the anomaly is consistent with the expected estimate in Eq. (\ref{aestimate}) as suggested by some theoretical models. It reaches a tenth of mm$/$s$^2$ at
some instants with an error estimate from Eq. (\ref{deltaerr}) below $0.001$ mm$/$s$^2$.
On the other hand, this effect seems to be closely related to the proximity of the spacecraft to the surface of the Earth as the three components of the anomalous acceleration rapidly diminish with the time before of after the perigee. 

Another interesting feature is that the sign of the radial acceleration changes as the spacecraft crosses its perigee. To check if this pattern persists in other flybys we have performed the same analysis for the first and second Galileo flybys, the
Cassini \cite{Anderson2008} and the Juno flybys \cite{Thompson,Acedo2015,IorioJuno}. The results are plotted in Figs. \ref{fig:8}-\ref{fig:10} where the change in the sign after perigee of the anomalous acceleration vector is clear for the radial, polar and azimuthal components. We have also modelled the Rosetta and Messenger flybys of 2005 but in these cases the anomalous acceleration is in the range $(-0.01,0.01)$ mm$/$s$^2$, so it cannot be discriminated from statistical fluctuations. This way, we confirm the absence or negligible value for the flyby anomaly for these flybys \cite{Anderson2008}. A first attempt to model this behaviour could be given by:
\begin{equation}
\label{amodel}
\delta a_i = \alpha_i \, g_0 \, e^{-h/L_i} \, F(\theta) \, \displaystyle\frac{\dot{r}}{c} \; , \;  i=r\, ,\theta\, ,\lambda\; ,
\end{equation}
where $i$ stands for the radial, polar or azimuthal component, $\dot{r}/c$ is the radial velocity ratio with the speed of light and $F(\theta)=\cos \theta$ for the radial component or $F(\theta)=\sin \theta$ for the polar or azimuthal ($\theta$ denotes the colatitude). Here $\alpha_i$, $L_i$ are a non dimensional constant and a length scale, respectively, $h$ is the 
distance of the spacecraft to the Earth's surface and $g_0=9.8$ m$/$s$^2$ is the surface gravity.
The fittings in Figs. \ref{fig:5}-\ref{fig:10} were obtained with $\alpha_r=-3$, $\alpha_\theta=1$, $\alpha_\lambda=-1.5$ and $L_i=1060$ km for $i=r$, $\theta$, $\phi$. The proposed interaction in Eq. \ref{amodel} would be a medium ranged fifth-force
proportional to the ratio $\dot{r}/c$ and, consequently, much larger that the corrections to Newtonian gravity provided by
General Relativity in the region of a few thousand kilometers around the Earth. The agreement with the data of this, very preliminary, approach is not good for every flyby and additional components of the anomalous force should, probably, be necessary. Anyway, we should refrain at the present status of the data analysis from further speculation as more data should be
collected in the future to validate these models.

On the other hand, we will integrate the proposed force term in Eq. (\ref{amodel}) along the hyperbolic
trajectory for the NEAR and the Cassini flybys to compare with the independent analysis at JPL, performed by Anderson et al. \cite{Anderson2008}, in which the flyby anomaly was discovered. This should provide a test of consistency for our method.
The anomalous acceleration vector is given by $\delta \mathbf{a}_{\mbox{pert}}=\delta a_r \hat{r}+\delta a_{\theta} \hat{\theta}+\delta a_{\lambda} \hat{\lambda}$. The integration along the trajectory is then:
\begin{equation}
\label{vpert}
\delta \mathbf{v}(T)=\displaystyle\int_0^T \, \delta \mathbf{a}_{\mbox{pert}} \, d t \; ,
\end{equation}
where $t$ is the time since the crossing of the perigee and $\delta \mathbf{v}(T)$ is the perturbation in the velocity at time
$T$ after, or before, the perigee.

Following Anderson et al. \cite{Anderson2008} we should define the variation of the velocity modulus at time $T$ with respect to the ideal hyperbolic trajectory as follows:
\begin{equation}
\delta V_{\mbox{pert}}= \left\vert \mathbf{V}+\delta \mathbf{v}(T) \right\vert-\left\vert \mathbf{V} \right\vert \simeq \displaystyle\frac{\mathbf{V}}{\left\vert \mathbf{V} \right\vert} \cdot \delta\mathbf{v}(T) \; ,
\end{equation}
where $\left\vert \mathbf{V} \right\vert$ denotes the modulus of the corresponding velocity vector in the keplerian ideal trajectory and the approximation, as the dot product of the vector in Eq. (\ref{vpert}) and Kepler's velocity vector, $\mathbf{V}$,  is obtained for small perturbations.

\begin{figure}
\includegraphics[width=\columnwidth]{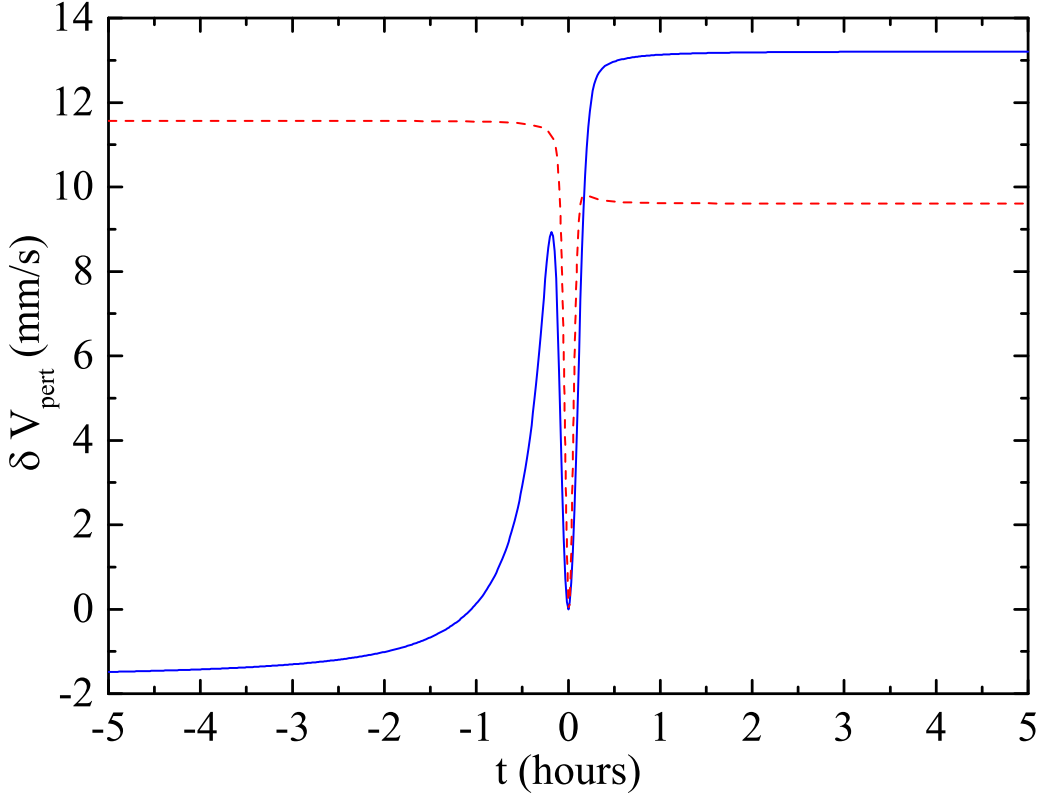}
\caption{The anomaly in the modulus of the velocity with respect to its value in the ideal keplerian orbit as a function of time
since the perigee in hours. Solid line denotes the case of the NEAR flyby and dashed line is for the Cassini flyby.}
\label{fig:11}       
\end{figure}

The results are plotted in Fig.\ (\ref{fig:11}) for the NEAR flyby with $\alpha_r=-3$, $\alpha_\theta=1$, $\alpha_\lambda=-1.5$
and, with these values of the parameters, we find a variation of the asymptotic post-encounter velocity with respect to the pre-encounter velocity of $14.70$ mm$/$sec. Maximum time for the integration was taken as $T=5$ hours after and before the perigee.
For the case of the Cassini flyby and with similar parameters ($\alpha_r=-2$, $\alpha_\theta=1.7$ and $\alpha_\phi=-0.8$) we obtain that the asymptotic velocity perturbation, with respect to the incoming asymptotic direction, is around $-1.96$ mm$/$s.
On the other hand, both values of $\delta V_{\mbox{pert}}$ are consistent with the previous orbital analysis \cite{Anderson2008}. The variability
in the coefficients of Eq. (\ref{amodel}) required to fit the velocity anomalies may indicate that these are not really constants but that an additional, more complicated, dependence with colatitude is present. We should highlight the relevance of this result to fundamental physics because it shows that, if the flyby anomaly is a real phenomenon, a force-field of the form given in Eq. (\ref{amodel}) maybe acting in a region close to the Earth's surface. This would be an effect of first order in the ratio of the spacecraft's velocity to the speed of light in contrast with the second order corrections predicted by
General Relativity.

\section{Conclusions}
\label{conclusions}

In this paper, we have discussed the development of an orbital model to analyze spacecraft flybys around the Earth in the
vicinity of the perigee. This model includes all known relevant contributions to the perturbations: tidal forces from the Sun and 
the Moon, atmospheric friction for low perigee flybys and zonal, tesseral and sectorial harmonics of the geopotential model. Other minor
effects have also been considered in the error analysis: ocean tides, corrections provided by General Relativity, perturbations
by other bodies in the Solar system such as Jupiter, errors in the zonal, tesseral and sectorial harmonic coefficients, determination of
celestial coordinates and terrestrial longitude and latitude and errors in the numerical procedure.

The objective of this study has been to extract reliable information about the perturbations in the position of the spacecraft
from approximately half an hour before the perigee to half an hour afterwards. And from this data, to derive the magnitude and
components of the anomalous acceleration acting upon the spacecraft, i. e., the acceleration that should be imparted upon the
spacecraft to obtain an exact agreement among the predicted and the observed trajectories
(By observed trajectories we mean the fittings to telemetry data incorporated in the Horizons database by the mission teams \cite{Horizons}. These trajectories take into account the information processing of the Doppler traking of each spacecraft \cite{Giorgini}). We have found that such acceleration
can be deduced from the numerical method and that its magnitude agrees with the expected estimate provided by some modified models of gravity \cite{Acedo2015,Acedo2017three} as given in Eq. (\ref{aestimate}). This acceleration peaks at values around
$0.1$ mm$/$s$^2$ and it is characterized by nonzero radial, polar and azimuthal components that, in most cases, change sign when
the spacecraft crosses its perigee. Moreover, this anomaly decreases very fast with altitude with a characteristic length scale around $1000$ km. These features are consistent with the existence of an unknown fifth field of force around the Earth beyond
standard physics. This field is more intense than the relativistic corrections to Newtonian gravity in that regime and we suggest that it could be proportional to the ratio among the spacecraft radial velocity and the speed of light. We should also mention that a phenomenological formula for an anomalous force decaying exponentially from the Earth surface was already considered
by H. J. Busack \cite{BusackI}. But Busack's formula, depending upon the spacecraft and Sun's velocity in a given reference frame, turned out to be incorrect when applied to the Juno flyby of the Earth in 2013 \cite{BusackII}. 

A fifth force of a different nature was considered in 80's of the past century but finally dismissed \cite{Franklin}. Recently, there have also been a proposal for a protophobic fifth force mediated by a new boson as an explanation of certain anomalies in
transitions of $^8$Be \cite{Feng}. In the case of larger scales there is also a possibility for a fifth force which should manifest itself
as a modification of standard gravity. As this force is expected to be proportional to the inertial mass, it would be, fundamentally, gravitational in origin but not considered in the formalism of general relativity. We must also point out that
the present status for the experimental verification of the General Relativity theory of gravity is not comparable with that
of the other interactions \cite{Will}. In addition, the form of the lagrangian in this case is not constrained by gauge and renormalizability conditions as those of the electroweak and strong interactions \cite{Peskin}.

In any case, an inductive scientific approach for the elucidation of this riddle would require further experimental data
as could have been provided by the STE-QUEST mission \cite{STEQUEST}, now cancelled. Research into the flyby anomaly is, consequently, in dire need of new reliable data and pursuing the theoretical analysis could help to plan
orbital analysis as a scientific mission objective in future missions. Another opportunity to the study of
the anomalies is provided by the recent Juno mission, in which the
spacecraft is planned to perform $36$ highly elliptical orbits of Jupiter with a perigee at roughly $4200$ km over the top
clouds of the planet. Although the gravitational field of Jupiter is not known with the detail of that of the Earth if would be highly interesting to analyze these orbits, with similar procedures as those described in this paper, to find if a similar
anomaly is found. In this study, the most recent determination of Jupiter's zonal harmonics from the first data sets of Juno's orbit can be helpful \cite{Bolton2017}. If the anomaly is discovered also in this case we will have an important science case. Moreover, we expect larger anomalous accelerations for the flyby of Jupiter by the Juno spacecraft ($\simeq 10$ times those found in the case of the Earth). Work along this line
is in progress and it will be published elsewhere.

%
%

%

%
\section*{Acknowledgements}

I gratefully acknowledge NASA's JPL  for providing their orbital fits, for the missions considered in this paper, through the Horizon's website.


%
 \bibliographystyle{hplain}  
 \bibliography{acedobiblio}                

%

\end{document}